\documentclass{elsart}

\usepackage{graphicx}
\usepackage{ifthen}
\usepackage{xspace}

\usepackage{amsmath}

\newcommand{\mevcc}{\ensuremath{\mathrm{MeV}/c^2}\xspace}

\newcommand{\gevc}{\ensuremath{\mathrm{GeV}/c}\xspace}
\newcommand{\gev}{\ensuremath{\mathrm{GeV}}\xspace}

\newcommand{\gevb}{\ensuremath{\mathrm{GeV}^{-2}c^2}\xspace}

\newcommand{\figlabel}[1]{\label{fig:#1}}

\begin{document}

\begin{frontmatter}
\title{\boldmath Observation of a 1750~\mevcc Enhancement in the Diffractive Photoproduction of $K^+K^-$}

\date{\today}

The FOCUS Collaboration\footnote{see http://www-focus.fnal.gov/authors.html for
additional author information.}
\author[ucd]{J.~M.~Link}
\author[ucd]{M.~Reyes}
\author[ucd]{P.~M.~Yager}
\author[cbpf]{J.~C.~Anjos}
\author[cbpf]{I.~Bediaga}
\author[cbpf]{C.~G\"obel}
\author[cbpf]{J.~Magnin}
\author[cbpf]{A.~Massafferri}
\author[cbpf]{J.~M.~de~Miranda}
\author[cbpf]{I.~M.~Pepe}
% ...

\address[ucd]{University of California, Davis, CA 95616}
\address[cbpf]{Centro Brasileiro de Pesquisas F\'isicas, Rio de Janeiro, RJ, Brasil}
\address[cinv]{CINVESTAV, 07000 M\'exico City, DF, Mexico}
\address[cu]{University of Colorado, Boulder, CO 80309}\nopagebreak
\address[fnal]{Fermi National Accelerator Laboratory, Batavia, IL 60510}
\address[fras]{Laboratori Nazionali di Frascati dell'INFN, Frascati, Italy I-00044}
\address[ui]{University of Illinois, Urbana-Champaign, IL 61801}
\address[iu]{Indiana University, Bloomington, IN 47405}
\address[korea]{Korea University, Seoul, Korea 136-701}
\address[kp]{Kyungpook National University, Taegu, Korea 702-701}
\address[milan]{INFN and University of Milano, Milano, Italy}
\address[nc]{University of North Carolina, Asheville, NC 28804}
\address[pavia]{Dipartimento di Fisica Nucleare e Teorica and INFN, Pavia, Italy}
\address[pr]{University of Puerto Rico, Mayaguez, PR 00681}
\address[sc]{University of South Carolina, Columbia, SC 29208}
\address[ut]{University of Tennessee, Knoxville, TN 37996}
\address[vu]{Vanderbilt University, Nashville, TN 37235}
\address[wisc]{University of Wisconsin, Madison, WI 53706}

 
\begin{abstract}
Using the FOCUS spectrometer with photon beam energies between 20 and 160~\gev, we confirm the existence of a diffractively photoproduced enhancement in $K^+K^-$ at 1750~\mevcc with nearly 100 times the statistics of previous experiments.  Assuming this enhancement to be a single resonance with a Breit-Wigner mass shape, we determine its mass to be $1753.5\pm 1.5\pm 2.3$~\mevcc and its width to be $122.2\pm 6.2\pm 8.0$~\mevcc.  We find no corresponding enhancement at 1750~\mevcc in $K^*K$, and again neglecting any possible interference effects we place limits on the ratio $\Gamma (X(1750) \rightarrow  K^*K)/\Gamma (X(1750) \rightarrow K^+K^-)$.  Our results are consistent with previous photoproduction experiments, but, because of the much greater statistics, challenge the common interpretation of this enhancement as the $\phi (1680)$ seen in $e^+e^-$ annihilation experiments.
 
\end{abstract}
\end{frontmatter}

\section{Motivation}

Previous photoproduction experiments have consistently observed an enhancement in $K^+K^-$ at a mass near 1750~\mevcc (referred to here as the X(1750))~\cite{Ast81,Atk85,Bus89}.  However, with signals consisting of only around 100 events and with a statistical significance of the enhancement of, at best, only $3.5~\sigma$ over background, these experiments have suffered from a lack of statistics.  Due to the large statistical errors on the mass of the enhancement, and assuming that a diffractively photoproduced meson is a vector, this enhancement has been identified with the $\phi (1680)$ seen in $e^+e^-$ annihilation~\cite{pdg,Cle94,Bis91,Buo82}, which is a candidate radial excitation of the $\phi (1020)$.  The present analysis challenges this interpretation of the 1750~\mevcc signal on two grounds.

First, assuming the enhancement is a single resonance, we determine a mass of $1753.5\pm 1.5\pm 2.3$~\mevcc from our sample of more than 10,000 signal events, which is clearly inconsistent with 1680~\mevcc.  One previous photoproduction experiment~\cite{Ast81}, using less than 50 signal events and guided by $e^+e^-$ annihilation results, presented a mass of $1690\pm 10$~\mevcc after an analysis based on a model including interference and Deck-like effects.  With orders of magnitude more statistics, we are unable to reproduce this result.

Second, although $e^+e^-$ annihilation experiments report the dominant decay mode of the $\phi (1680)$ to be $K^*K$~\cite{Buo82}, as expected theoretically for the radial excitation of the $\phi (1020)$~\cite{God85,God95a,God95b}, we find no evidence for a photoproduced enhancement in $K^*K$ corresponding to the photoproduced enhancement in $K^+K^-$.  We put a low upper limit on the ratio $\Gamma (X(1750) \rightarrow  K^*K)/\Gamma (X(1750)\\ \rightarrow K^+K^-)$.

\section{Data Sample}

The data for this analysis was collected by the Wideband photoproduction experiment FOCUS during the Fermilab 1996--1997 fixed-target run.  FOCUS is a considerably upgraded version of the previous E687 photoproduction experiment~\cite{Fab92}.  A forward multi-particle spectrometer is used to measure the interactions of high energy photons on a segmented BeO target.

The FOCUS detector is a large aperture, fixed-target spectrometer with excellent vertexing, particle identification~\cite{xxxid}, and reconstruction capabilities for photons, $\pi^0$'s, and $K_S$~\cite{xxxks}.  A photon beam is derived from the bremsstrahlung of secondary electrons and positrons with an approximately 300~\gev endpoint energy produced from the 800~\gev Tevatron proton beam.  The charged particles which emerge from the target are tracked by two systems of silicon microvertex detectors.  The upstream system, consisting of 4 planes (two views in 2 stations), is interleaved with the experimental target, while the other system lies downstream of the target and consists of twelve planes of microstrips arranged in three views.  The momentum of a charged particle is determined by measuring its deflections in two analysis magnets of opposite polarity with five stations of multiwire proportional chambers.  Three multicell threshold Cerenkov counters are used to discriminate between pions, kaons, and protons.

The $K^+K^-$ and $K_SK^{\pm}\pi^{\mp}$ data samples used in this paper require a single vertex in the target, no electromagnetic energy apart from that associated with the reconstructed tracks, and require all tracks to be singly linked between the upstream and downstream tracking systems.  Events with additional reconstructed tracks are rejected, making the data samples as exclusive as possible.  The Cerenkov identification of the kaons in both diffractive final states limits the photon energy range to $\leq 160$~\gev. 

\section{Mass and Width Measurement}

Our sample of $K^+K^-$ events, selected using the criteria described above, shows a clear $\phi (1020)$ signal dominating the spectrum (Figure 1).  The diffractive component of the production of the $\phi (1020)$ shows up as a peak in the $p_T$ spectrum (Figure 2.a).  Cutting around this peak by requiring $p_T < {\rm 0.15}$~\gevc, we select a diffractive sample of $K^+K^-$ events, in which a clear enhancement appears in the mass spectrum near 1750~\mevcc (Figure 3.a).  Figure 3.b confirms that the enhancement appears at only low $p_T$.  Plotting the $p_T$ spectra in the 1750 region (1640--1860~\mevcc) and in the two sideband regions (1500--1600~\mevcc and 1900--2100~\mevcc), it is seen that the 1750 region has a peak in the $p_T$ spectrum in nearly the same place as the $\phi(1020)$ peak, but the sideband regions have significantly smaller $p_T$ peaks (Figure 2.a), indicating that the background under the X(1750) signal is largely non-diffractive.

\begin{figure}
 \centering
  \includegraphics[width=6.4cm]{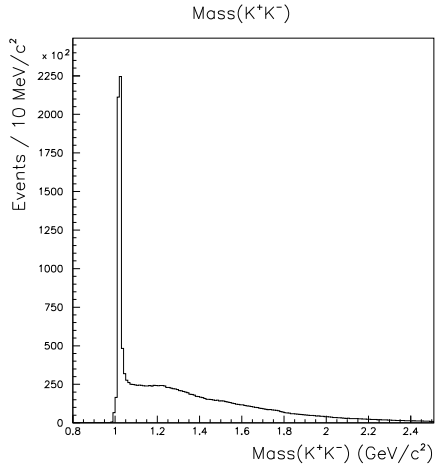}
 \caption{The $K^+K^-$ sample with no cut on $p_T$.}
 \figlabel{fig_1}
\end{figure}

The $t'$ spectrum ($t' \equiv |t| - |t|_{min} \approx p_T^2$ and $t \equiv (P_{\gamma} - P_{KK})^2$, where $P_{\gamma}$ and $P_{KK}$ are the four-momenta of the photon beam and the KK system, respectively) for $K^+K^-$ masses between 1640 and 1860~\mevcc has been fit with two exponentials (Figure 2.b).  The diffractive exponential for this region has a slope of $69.2\pm 2.1$~\gevb and the background exponential has a slope of $4.17\pm 0.21$~\gevb.  For comparison, the $\phi (1020)$ signal has a diffractive slope of $77.71\pm 0.59$~\gevb and a background slope of $1.71\pm 0.14$~\gevb.  The steep diffractive slopes are characteristic of exclusive diffractive photoproduction off of nuclear targets.

\begin{figure}
 \centering
  \includegraphics[width=10cm]{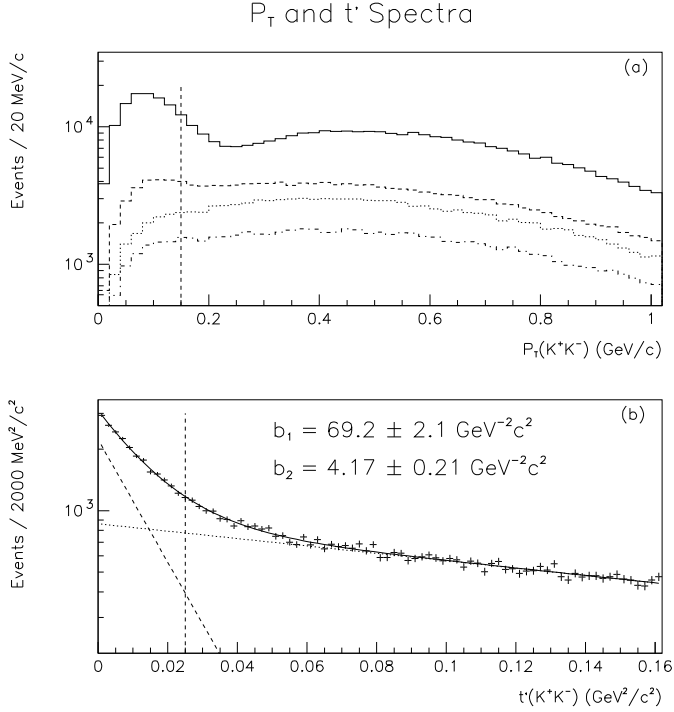}
 \caption{(a) The $K^+K^-$ $p_T$ spectra.  The solid line is the $p_T$ spectrum for the $\phi (1020)$.  The top dotted line is the $p_T$ spectrum for $K^+K^-$ masses between 1640 and 1860~\mevcc; the middle is the left sideband (1500--1600~\mevcc); and the bottom is the right sideband (1900--2100~\mevcc).  (b) The $t'$ spectrum for $K^+K^-$ masses between 1640 and 1860~\mevcc fitted using two exponentials, $A_1e^{-b_1t'}+A_2e^{-b_2t'}$.  In each plot, the vertical line represents the $p_T$ cut used in this analysis.}
 \figlabel{fig_2}
\end{figure}

Fitting the 1750~\mevcc mass region with a non-relativistic Breit-Wigner distribution and a quadratic background, we find

\begin{displaymath}
\textrm{Yield} = 11,700\pm 480~{\rm Events}
\end{displaymath}
\begin{displaymath}
\mathrm{M} = 1753.5\pm 1.5\pm 2.3~\mevcc
\end{displaymath}
\begin{displaymath}
\mathrm{\Gamma} = 122.2\pm 6.2\pm 8.0~\mevcc
\end{displaymath}

Because the acceptance of the detector is flat in this region, as determined by a full Monte Carlo simulation, the fit was performed on the uncorrected mass spectrum.  Further, since Monte Carlo studies of the detector have shown that the $K^+K^-$ mass resolution in the 1750~\mevcc region is better than 10~\mevcc, which is much less than the width of the X(1750), resolution effects have been neglected in the fit.  (Similar Monte Carlo studies have accurately predicted the $D^0$ mass resolution in the decay $D^0 \rightarrow K^+K^-$, and find a compatible resolution~\cite{xxxDs}.)  The systematic errors were determined by varying the $p_T$ cut, the Cerenkov cuts, the form of the Breit-Wigner shape (non-relativistic, and relativistic L = 0,1,2), and the form of the background shape, and include the systematic uncertainty in our mass scale.

\begin{figure}
 \centering
  \includegraphics[width=10cm]{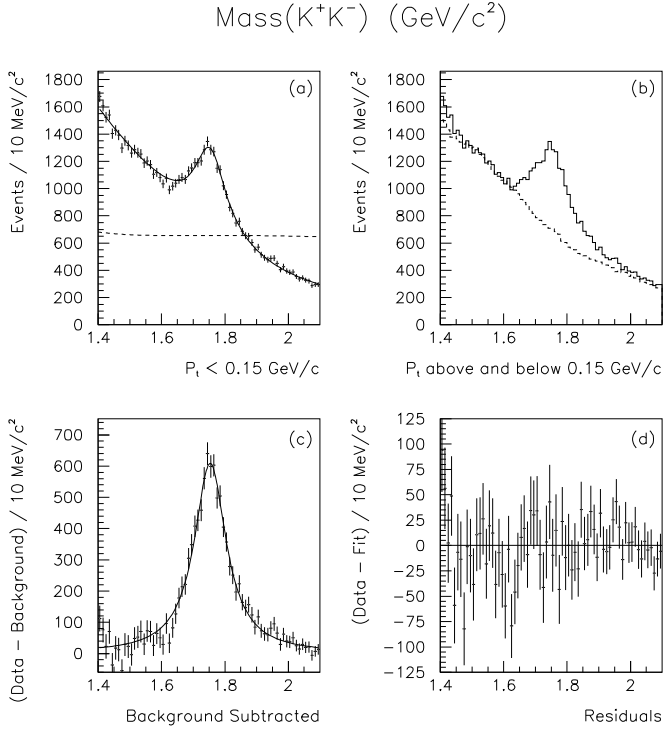}
 \caption{(a) The $K^+K^-$ mass spectrum with the requirement that $p_T < {\rm 0.15}$~\gevc.  The spectrum is fit with a non-relativistic Breit-Wigner distribution and a quadratic background.  The dotted line is the Monte Carlo efficiency on a scale from 0 to 100${\rm \%}$.  (b) The solid line is the $K^+K^-$ mass spectrum with the requirement that $p_T < {\rm 0.15}$~\gevc.  The dotted line is the $K^+K^-$ mass spectrum with $p_T > {\rm 0.15}$~\gevc scaled to the size of the low $p_T$ spectrum for comparison.  (c) The data and fit after subtracting the quadratic polynomial background shape.  (d) The data minus the fit.}
 \figlabel{fig_3}
\end{figure}

There is a region near 1600~\mevcc where there is some discrepancy in our fit to the $K^+K^-$ mass spectrum.  The residuals show that the statistical significance of this discrepancy is not strong (Figure 3.d).  It has been found that several different interference scenarios can improve the fit.  These include interference with the $K^+K^-$ continuum and interference between the X(1750) and a second resonance with lower mass.  The goodness of the fits, however, does not allow us to discriminate between solutions, and we find no physics motivation for picking one solution over another.  In all scenarios, the mass of the X(1750) exceeds 1747~\mevcc.  If the X(1750) has $J^{PC} = 1^{--}$ and if the $\phi (1680)$ were also photoproduced, then we would expect some distortion of the $K^+K^-$ mass shape.  However, the apparent absence of the $\phi (1680)$ in its dominant $K^*K$ mode and the measured ratio of branching fractions~\cite{Buo82} limit the amplitude in the $K^+K^-$ state so that the distortion of the expected line shape is negligible.  Our mass and width measurements and our determination of systematic errors assume the production of a single, non-interfering resonance.

\section{Branching Ratio}

Figure 4 shows our $K_SK\pi$ sample and the two $K^*$ combinations.  The $K_SK^{\pm}\pi^{\mp}$ spectrum shows the classic ``D'' and ``E'' regions~\cite{God99}.

\begin{figure}
 \centering
  \includegraphics[width=10cm]{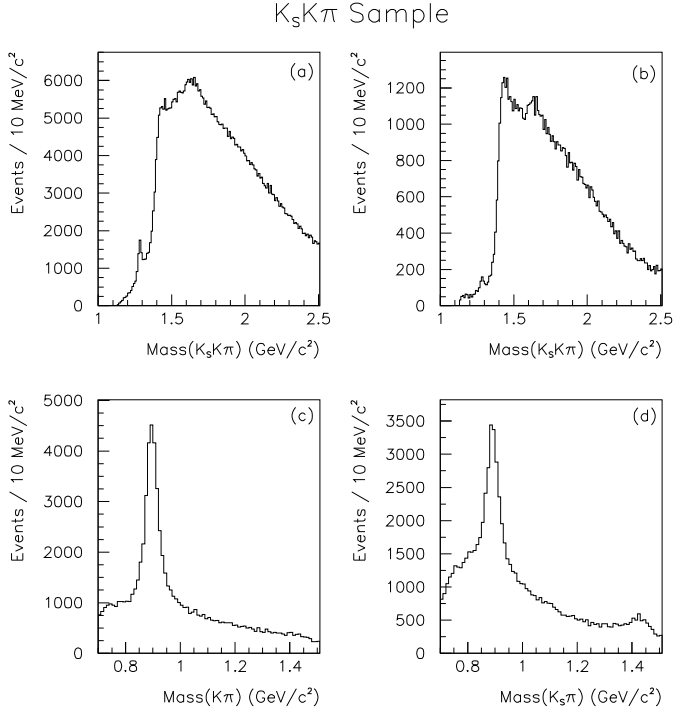}
 \caption{(a) The $K_SK\pi$ sample with no cut on $p_T$.  (b) The $K_SK\pi$ sample with $p_T < {\rm 0.15}$~\gevc.  (c) The $K_S\pi^{\pm}$ mass from (b).  (d) The $K^{\pm}\pi^{\mp}$ mass from (b).}
 \figlabel{fig_4}
\end{figure}

After requiring $p_T < {\rm 0.15}$~\gevc (the same $p_T$ cut imposed on the $K^+K^-$ sample), and requiring a $K^*$, the two distinct $K^*K$ spectra were fit individually (Figure 5).  There is no obvious X(1750) signal in either of the two $K^*K$ modes.  In order to place upper limits on the $K^*K/K^+K^-$ branching ratios of the X(1750), an estimate of the background is needed.  The presence of a slight enhancement somewhat below the $\phi (1680)$ region introduces some ambiguity in estimating the background in the X(1750) region.  In order to make a conservative estimate of the background, we have used a fit which includes a second, unconstrained, non-interfering resonance in the 1630~\mevcc region as well as the X(1750) with mass and width fixed from the fit to the $K^+K^-$ mode.  The size of the X(1750) resonant component was unconstrained.  These fits provide an estimate of the number of events above background in the X(1750) region, $-123 \pm 120$ events in the $K^{*0}K_S$ mode and $106 \pm 117$ in the $K^{*\pm}K^{\mp}$ mode.

\begin{figure}
 \centering
  \includegraphics[width=10cm]{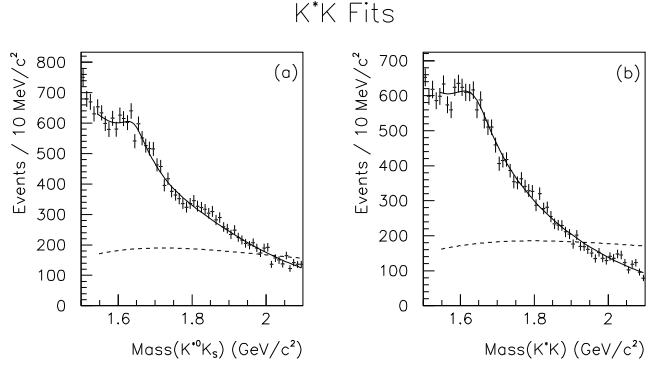}
 \caption{Fits to $K^*K$ using a Breit-Wigner distribution for the X(1750) region with mass and width fixed from the fit to $K^+K^-$, a second non-interfering Breit-Wigner distribution around 1630~\mevcc, and a quadratic background.  The dotted lines are the Monte Carlo efficiencies on a scale from 0 to 20${\rm \%}$.  (a) $K^{*0}K_S$ with $K^{*0}$ to $K^{\pm}\pi^{\mp}$.  (b) $K^{*\pm}K^{\mp}$ with $K^{*\pm}$ to $K_S\pi^{\pm}$.}
 \figlabel{fig_5}
\end{figure}

The efficiencies of the $K^+K^-$ and $K^*K$ final states were determined by Monte Carlo simulations in bins of $p_T$, beam energy, and mass.  As the spin of the X(1750) is still uncertain, several decay angular distributions were simulated.  Using the highest efficiency for $K^+K^-$ (Figure 3.a) and the lowest for $K^*K$ (Figure 5) and correcting for the $K_S$ unseen decay mode, we have found an upper limit on the following relative branching ratios

\begin{displaymath}
\frac{\Gamma (X(1750) \rightarrow  \overline{K}^{*0}K^0 \rightarrow K^{-}\pi^{+}K_S+c.c.)}{\Gamma (X(1750) \rightarrow K^+K^-)} <0.065~{\rm at}~90 \% ~{\rm C.L.} 
\end{displaymath}
\begin{displaymath}
\frac{\Gamma (X(1750) \rightarrow  K^{*+}K^{-} \rightarrow K_S\pi^{+}K^{-}+c.c.)}{\Gamma (X(1750) \rightarrow K^+K^-)} <0.183~{\rm at}~90 \% ~{\rm C.L.} 
\end{displaymath}

The confidence limits were set using the Feldman-Cousins methodology~\cite{Fel98}.  The two relative branching ratios were measured to be $-0.083 \pm 0.081$ and $0.065 \pm 0.072$, respectively.

\section{Discussion}

Because of the large discrepancies in mass and relative branching fractions to $K^+K^-$ and $K^*K$, we do not believe it is reasonable to identify the X(1750) with the $\phi (1680)$.  In fact, because the mass of the X(1750) is significantly higher than all known vector mesons, the most massive of which are the $\omega (1650)$, $\phi (1680)$, and $\rho (1700)$, an interpretation claiming the X(1750) is some combination of interfering vector mesons also seems highly unlikely.  The interpretation of the X(1750) remains uncertain.

\section{Acknowledgements}

We wish to acknowledge P. Page and T. Barnes for valuable discussions, the assistance of the staffs of Fermi National Accelerator Laboratory, the INFN of Italy, and the physics departments of the collaborating institutions. This research was supported in part by the U.~S. National Science Foundation, the U.~S. Department of Energy, the Italian Istituto Nazionale di Fisica Nucleare and Ministero dell'Universit\`a e della Ricerca Scientifica e Tecnologica, the Brazilian Conselho Nacional de Desenvolvimento Cient\'{\i}fico e Tecnol\'ogico, CONACyT-M\'exico, the Korean Ministry of Education, and the Korean Science and Engineering Foundation.

\end{document}